\documentclass[journal]{IEEEtran}
\usepackage{cite}

\ifCLASSINFOpdf
   \usepackage[pdftex]{graphicx}
\else
\fi
\usepackage{amsmath}

\usepackage{color}

\title{Perceived Audiovisual Quality Modelling based on Decison Trees, Genetic Programming and Neural Networks}

\author{Edip~Demirbilek, 
Jean-Charles Gr{\'e}goire
\thanks{This work is supported by NSERC and Summit-Tech Inc., under the Collaborative Research and Development grant programme.}
\thanks{E. Demirbilek and JC. Gr{\'e}goire are with Institut National de la Recherche Scientifique (INRS), Montr{\'e}al, CANADA}
}

\begin{document}

\maketitle

\begin{abstract}

Our objective is to build machine learning based models that predict audiovisual quality directly from a set of correlated parameters that are extracted from a target quality dataset. We have used the bitstream version of the INRS audiovisual quality dataset that reflects contemporary real-time configurations for video frame rate, video quantization, noise reduction parameters and network packet loss rate. We have utilized this dataset to build bitstream perceived quality estimation models based on the Random Forests, Bagging, Deep Learning and Genetic Programming methods.

We have taken an empirical approach and have generated models varying from very simple to the most complex depending on the number of features used from the quality dataset. Random Forests and Bagging models have overall generated the most accurate results in terms of RMSE and Pearson correlation coefficient values. Deep Learning and Genetic Programming based bitstream models have also achieved good results but that high performance was observed only with a limited range of features. We have also obtained the epsilon-insensitive RMSE values for each model and have computed the significance of the difference between the correlation coefficients.

Overall we conclude that computing the bitstream information is worth the effort it takes to generate and helps to build more accurate models for real-time communications. However, it is useful only for the deployment of the right algorithms with the carefully selected subset of the features. The dataset and tools that have been developed during this research are publicly available for research and development purposes.
\end{abstract}


%


\section{Introduction}
\label{sec:intro}

\begin{figure*}[ht!]
  \centering
    \includegraphics[width=0.8\textwidth]{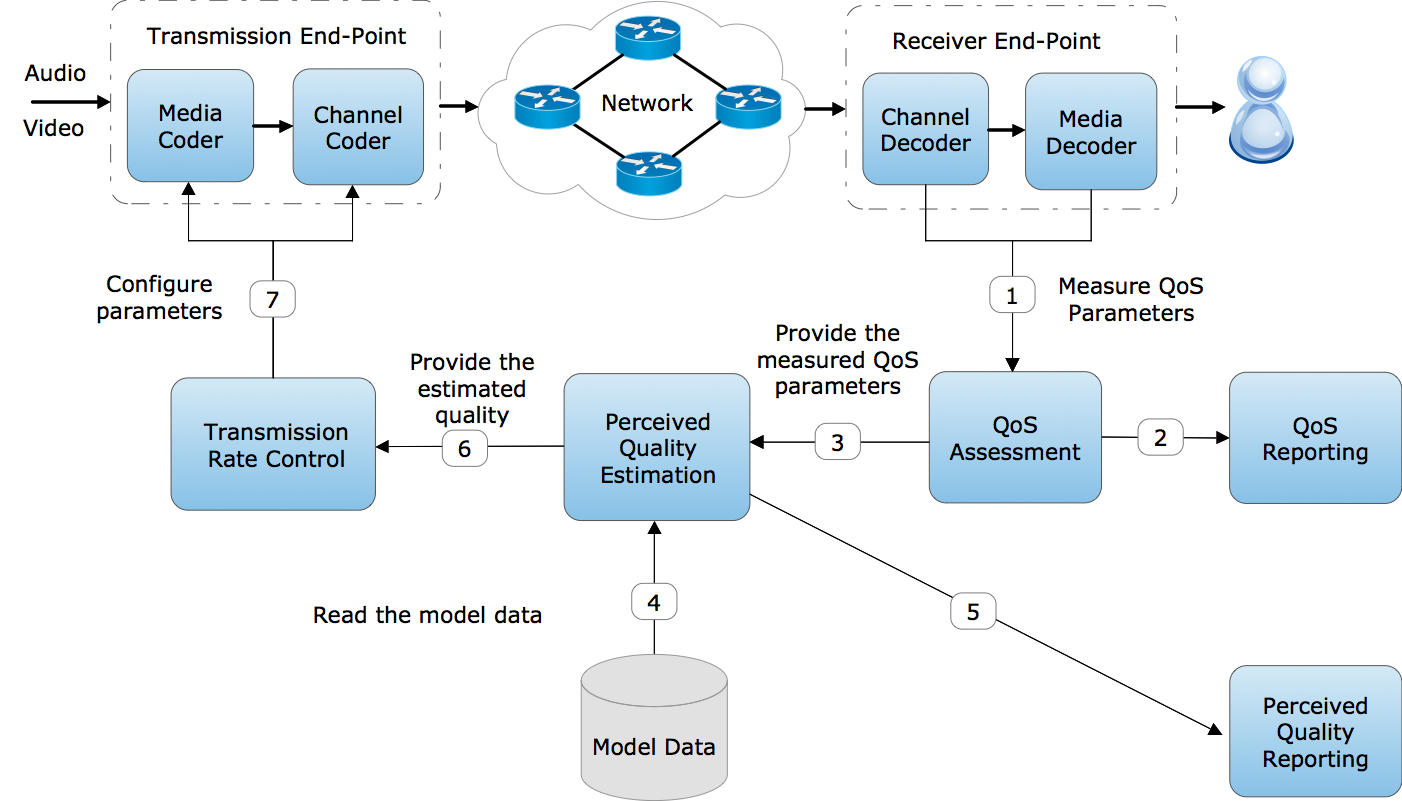}
    \caption{Sample Architecture: Perceived quality modeling and perceived quality based multimedia quality adaptation for real-time communications.}
    \label{fig:QoEFeedback}
\end{figure*}

One of the standard quality assessment methods of multimedia communications is called Quality of Service (QoS) and is based on the measure of parameters such as throughput, delay, jitter and packet loss rate \cite{perkiselectronic}. When analyzing the characteristics of a real-time multimedia communication, it is necessary to have a clear understanding of changes in these parameters. This is of particular importance in environments where the multimedia streams are traversing resources that have limited and fluctuating bandwidth capacity, such as wireless networks \cite{hansen2013assessing}.

In order to cope with ever-changing situations during a multi-party multimedia interaction, one of the standard approaches is to periodically measure various QoS parameters and feed them back to the transmission endpoint in order to control the rate and characteristics of the streamed multimedia. Such an adaptation model generally improves the quality of audiovisual communications. These QoS measurements can also be reported for quality monitoring.

There has been significant research about the QoS feedback control in both application and network layers and various management tools and approaches have been developed \cite{bordetsky2001feedback}. However, in recent years, the convergence of the digital media industry and the information and communications technology (ICT) industry has led to a paradigm shift away from QoS towards perceived multimedia presentation quality \cite{perkiselectronic}.

In perceived quality based multimedia adaptation, although a better QoS usually provides a higher presentation quality, there are cases where QoS trade-offs are needed to maximize the overall perceived quality \cite{hansen2013assessing}. These trade-offs could be made statically and dynamically for a real-time communication.

Figure \ref{fig:QoEFeedback} depicts a sample architecture of a multimedia communication system where QoS parameters are measured at regular intervals and reported for monitoring as well as for parameter tuning to the transmission endpoint along with the perceived quality predictions. Here, audio and video data streams are generated and encoded on the transmission endpoint and then forwarded to the network. On the receiver endpoint, the data stream is first decoded at the channel level and then at the media level, and finally presented to the end-user in conventional ways. While this process takes place, in a closed loop, these measured QoS parameter values and the quality predictions are evaluated by the transmission rate control, and media and channel encoding parameters modified to improve the user experience. Service Providers can enhance their service monitoring and reporting processes dramatically by exploiting such quality predictions. This sample architecture assumes that perceived quality modeling is already implemented properly and this is the main concern of this paper.

The classic approach to audiovisual quality modeling is to develop functions to predict audio and video quality independently and then combine them using another function to predict the overall audiovisual perceived quality. The alternative way is to build models that predict the audiovisual quality directly without any intermediate functions. Machine learning based modeling approaches have been successfully applied to estimating perceived quality \cite{aroussi2014survey} \cite{maki2013reduced}. In this research, we have taken the second approach and have built machine learning based models that predict the overall audiovisual quality in one single function.

The rest of this paper is organized as follows. In sections \ref{sec:quality}, \ref{sec:Standardized} and \ref{sec:ml}, we present some background information on various aspects of the audiovisual quality modeling and the machine learning algorithms we have used. The bitstream version of the INRS audiovisual quality dataset is introduced in Section \ref{sec:dataset}. Reduced-reference bitstream audiovisual quality prediction models are presented in details in Section \ref{sec:bitstreammodels}. We sum up the work in the conclusion section.

\section{Perceived Quality Modelling}
\label{sec:quality}

The European Network on Quality of Experience in Multimedia Systems and Services (Qualinet) whitepaper \cite{le2012qualinet} defines Quality of Experience (QoE) as follows:
\textit{Quality of Experience (QoE) is the degree of delight or annoyance of the user of an application or service. It results from the fulfillment of his or her expectations with respect to the utility and/or enjoyment of the application or service in the light of the user’s personality and current state.}

Additionally, it defines the characteristics of a user, service, application, or context whose actual state or setting may have an influence on the QoE. These influence factors may be interrelated but still can be broadly classified into three categories, namely Human Influence Factors (IF), System Influence Factors, and Context Influence Factors.

Human IFs are complex and strongly interrelated and related to demographic and socio-economic background, physical and mental constitution, or emotional state \cite{le2012qualinet}.

System IFs describe the properties and characteristics that determine the technically produced quality of an application or service. These factors are related to media capture, coding, transmission, storage, rendering, and reproduction/display, as well as to the communication of information itself—from content production to user—and can be further sub-grouped into 4 categories: (1) Content-related System IFs, (2) Media-related System IFs, (3) Network-related System IFs and (4) Device-related System IFs \cite{le2012qualinet}.

Context IFs are related to situational properties that describe the users' environment in terms of physical, temporal, social, economic, task, and technical characteristics \cite{le2012qualinet}.

The definition for QoE and its characteristics described by the Qualinet whitepaper reflects the most recent standardized work in the literature. There have been many attempts to model, predict or find a correlation between QoE and IFs. Most authors concentrate on network-oriented QoS parameters such as Packet Loss Rate, Jitter, Delay, and Throughput that are a subset of the System IFs \cite{maia2014concise} \cite{paudyal2014study}. In this research, we will also focus on System IFs and therefore use the term ``perceived quality'' rather than ``Quality of Experience'' as the IFs we analyze are only a subset of all potential IFs.

Collecting the opinion of a set of users via subjective experiments is one of the standard ways of measuring the perceived quality of telecommunication systems. However, conducting such experiments is costly and time-consuming, and harder to perform for real-time communications \cite{li2014real}, which is the focus of our research. An alternative approach is to conduct these experiments once to collect quality scores and then develop instrumental methods to predict the mean of the users' perception of service quality.

Predicting the perceived quality from objective measurements provides only partial solutions. However, these partial solutions do provide insight on how perceived quality is affected by network QoS parameters. It is possible to correlate the QoS parameters with the measured perceived quality metrics and build an effective perceived quality-aware QoS model \cite{rifai2011brief}.

In the following sections, we will revisit this topic in more detail when discussing the standardized perceived quality estimation models. As we see from published approaches, quality assessment systems that exploit machine learning paradigms very promisingly attain a high degree of prediction accuracy \cite{aroussi2014survey}. Machine learning based approaches provide a theoretical and methodological framework to quantify the relationship between perceived quality and network QoS.

\subsection{Quality Models}
\label{subsec:QualityModels}

Audiovisual quality relies on audio and video quality and their interactions. The overall audiovisual quality can be estimated via a function directly regardless of how degradations affect audio and video quality individually. An alternative approach is to predict intermediate audio and video quality individually and then integrate these into an overall audiovisual quality. Raake et al. \cite{raake2011ip} suggest that a complex model that uses the intermediate audio and video functions would generate more accurate predictions.

There are different approaches in building audiovisual quality models. Raake et al. \cite{raake2011ip} have categorized the models based on the type of data they use.

\begin{itemize}

\item Parametric quality models predict the impact of encoding configurations and network impairments on multimedia quality. They typically use information extracted from packet headers and have no access to the packet payload data. These methods are well suited to the cases where the payload data is encrypted \cite{dubin2016real}.

\item Planning models are similar to the parametric models but differ on where the input information is acquired from. Planning models are based on service information available during the planning phase while parametric models take the input information acquired from an existing service.

\item Media or signal-based models include aspects of human perception and assess the physical characteristics of the dispatched signal. They utilize the decoded signal as input to calculate a quality score.

\item Bitstream based models exploit information from the elementary stream. These models typically process both the headers and the payload of the video bitstream. They process the bitstream header to extract transport-related information such as Transport Stream (TS) and/or Realtime Transport Protocol (RTP) time stamps and sequence numbers for packet loss detection. They process the payload of the video bitstream to extract a number of features such as picture type, number of slices, Quantization Parameter (QP), motion vector, type of each macroblock (MB) and its partitions, and the transform coefficients of the prediction residual.

\item Hybrid quality assessment models exploit information from packet headers, the elementary stream and the reconstructed pictures. The information for the reconstructed pictures is obtained from the processed video sequence generated by an external decoder rather than from an internal decoder within the model.

\end{itemize}

Quality models can also be grouped relying on the type of additional information they process. Full-Reference (FR) models typically process the original source sequence while Reduced-Reference (RR) models use only a limited amount of information derived from the source sequence. No-Reference (NR) models use transmitted sequences without using any information from the original signal.

One important aspect of perceived quality modeling is that an objective model is not expected to predict an average subjective opinion more accurately than an average test subject. The uncertainty of the subjective votes is calculated via standard deviation and its corresponding Confidence Interval (CI) \cite{itut2012P.1401}.

\subsection{Statistical Metrics}
\label{subsec:Metrics}

Traditionally, the performance of a model is evaluated using three statistical metrics which are used to report the model performance's accuracy, consistency, and linearity/monotonicity \cite{itut2012P.1401}.

The accuracy of a model is usually determined by a statistical interpretation of the difference between the MOS values of the subjective test and its prediction on a generalized scale. An accurate model aims to make predictions with the lowest error in terms of ``Root Mean Square Error'' (RMSE) used during the subjective tests \cite{itut2012P.1401} \cite{garcia2014parametric}. ITU-T P.1401 suggests to compute the epsilon-insensitive RMSE (RMSE$^*$) to take the uncertainty of the subjective scores as well.

The perceived quality predictions have to have consistently low error margins over the range of test subjects. The model's consistency is reported by calculating either the residual error distribution or outlier ratio. Computing the outlier ratio requires finding the outliers that are determined as the points for which prediction error surpasses the 95\% confidence interval \cite{itut2012P.1401} \cite{garcia2014parametric}.

In the literature, there are two commonly used metrics for computing the linearity of a model: the Spearman rank coefficient and the Pearson correlation coefficient. The Pearson correlation coefficient is used whenever the sampled data has a near-normal distribution. In other cases, the Spearman rank coefficient is utilized to qualify the linearity between the predicted and the actual subjective quality scores \cite{garcia2014parametric}. ITU-T P.1401 additionally recommends to compute the significance of the difference between correlation coefficients using the  Fisher-z transformation which is determined for each comparison and evaluated against the 95\% t-Student value.

\subsection{Cross Validation}
\label{subsec:CrossVal}

Quality model predictions are compared to the actual quality scores to evaluate the performance of the models. However, in the case of limited amount of training and test data set, K-fold or ``leave-one-out cross-validation'' is used to report the performance of the quality model.

In the K-fold approach, the available data is split into K mutually exclusive subsets. In each step, K-1 subsets are used to train the data and the single remaining subset is used to measure the accuracy of the model. This procedure is repeated K times using a different portion of the available data as test data. Data splitting can be done randomly as well as by stratifying the subsets. Stratification process often improves the representativeness of the selected subset. Depending on the selected K subsets, the outcome of the model might vary. In order to make the predictions more robust and independent of the selected K subsets, it is recommended to repeat the procedure several times and then take the average of these runs for each metric. In this research, we have followed Kohavi's \cite{kohavi1995study} recommendation of using stratified ten-fold cross-validation for model selection.

In the leave-one-out cross-validation, K is equaled to the number of data points in the dataset and in each iteration, one data point is used to measure the accuracy of the model while all other data points are used to train the model \cite{kohavi1995study}.

\section{Standardized Audiovisual Quality Prediction Models}
\label{sec:Standardized}

Some of the standardized methods have been created through conducting competitions and selecting the models that have achieved the highest prediction accuracy. In this section, three audiovisual quality prediction models for streaming services and video telephony applications are briefly explained.

The ITU-T G.1071 \cite{itut2015G.1071} model is recommended for network planning of audio and video streaming services. This recommendation addresses higher resolution application areas like IPTV and lower resolution application areas like mobile TV. The application of the models is limited to QoE/QoS planning, and quality benchmarking and monitoring is outside the scope of this recommendation. The model takes network planning assumptions such as video resolution, audio, and video codec types and profiles, audio and video bitrates, packet-loss rate and distribution as input. As for output, it provides the separate predictions of audio, video and audiovisual quality defined on the five-point MOS scale. Use cases such as re-buffering degradation of audio, and video, transcoding situations, the effects of audio level, noise and delay, audiovisual streaming with significant rate adaptation are not covered by the model. It has been tested for low-resolution areas with ITU-T P.1201.1 training and validation databases and has achieved 0.50 RMSE and 0.83 Pearson correlation for audiovisual quality estimation. For high-resolution areas, it was tested on the ITU-T P.1201.2 training and validation databases and has achieved 0.51 RMSE and 0.87 Pearson correlation for audiovisual quality estimation.

The ITU-T G.1070 \cite{itut2012G.1070} is a planning model recommended for video telephony. In this model, overall multimedia quality is computed from network and application parameters as well as terminal equipment parameters. It proposes an algorithm that estimates videophone quality for the quality of experience and quality of service planners. It provides estimates of multimedia quality that take interactivity into account to allow planners to avoid under-engineering. The model contains three main functions for assessing speech quality, video quality, and overall multimedia quality. The speech quality estimation function is similar to the E-model \cite{rec2003g} and takes speech codec type, packet loss rate, bit rate, and talker echo loudness rating as input parameters. The video function is generated for “head-and-shoulders” content and takes video format, display size and codec type, packet loss rate, bit rate, key frame interval and frame rate as input parameters. The multimedia function integrates video alone and speech alone quality measures by including the audiovisual asynchrony and the end-to-end delay. The accuracy of the multimedia communication quality assessment model in terms of Pearson correlation was 0.83 for QVGA and 0.91 for QQVGA resolution on the given datasets. Application of the model is limited to QoE and QoS planning and other applications such as quality benchmarking and monitoring are not covered by the recommendation.

The ITU-T P.1201 \cite{itut2012P.1201} model is intended for estimating the audiovisual quality of streaming services. It is a non-intrusive packet-header information based model for the service monitoring and benchmarking of User Datagram Protocol (UDP) based streaming.  The model supports both lower resolution applications such as mobile TV and higher resolution applications such as Internet Protocol television (IPTV). It uses the information retrieved from the packet header as well as information provided out of band. It provides separate predictions of audio, video, and audiovisual quality as output in terms of the five-point MOS scale. It has been validated for compression, packet loss and re-buffering impairments of audio and video with different bitrates. Video content of different spatiotemporal complexity with different keyframes, frame rates, and video resolutions is selected. It was tested over 1166 samples at lower resolutions and tested over 3190 samples at higher resolutions. RMSE and Pearson correlation \cite{garcia2014parametric} values for audiovisual modeling were evaluated as 0.470 and 0.852, respectively for lower resolution applications and 0.435 and 0.911, respectively for higher resolution applications. Detailed performance figures are included in  \cite{itut2012P.1201}.

The ITU-T P.1201 model recommended for higher resolution applications \cite{itut2012P.1201.2} is a combination of the quality impairment based models where the quality model is based on the audio and video quality terms, and the impairment model is based on the audio and video impairment factor terms linked to the degree of compression and the transmission errors. The overall model takes the following mathematical form:

\begin{equation}
\label{eq:p12012}
Q_{AV} = \omega_{1} .Q_{AV} + \omega_{2} .IF_{AV}
\end{equation}

where \(\omega_{1}\) and \(\omega_{2}\) are constants, \(Q_{AV}\) denotes quality model and \(IF_{AV}\) denotes the impairment model.

Both quality and impairment models consist of complex mathematical forms that can be found in \cite{itut2012P.1201.2}. The impairment model requires the calculation of averaged number of bits per pixel (BitPerPixel) and scene-complexity (SceneComp) parameters. The scene-complexity parameter is calculated using video resolution, video frame rate, the number of scenes in the video sequence and the number of GOPs in the scene \cite{itut2012P.1201.2}. The implementation of the calculation of these parameters is given in the partial implementation of the ITU-T P.1201.2 audiovisual quality estimation tool in \cite{itut2012P.1201.2_tool}. We have integrated the BitPerPixel and SceneComp parameter evaluations from this tool into the generation of the bitstream version of the INRS audiovisual quality dataset.

ITU-T Rec. P.1201, G.1070, and  G.1071 have achieved high prediction accuracy against the respective test datasets provided. However, these methods have by definition limited application areas and cover limited coding technologies. Therefore, researchers have been attempting to improve these models ever since \cite{garcia2014parametric} \cite{belmudez2015audiovisual}.

In standardized models,  audio, video, and audiovisual predictions are performed by their respective functions whose output is then forwarded to the audiovisual function to predict overall audiovisual quality. An alternative approach is to implement the audiovisual function in a way that would not require intermediate predictions for audio and video quality and still be able to capture all those complex interrelations between influence factors. Machine learning based techniques have been successfully applied in implementing these functions \cite{maki2013reduced} \cite{gastaldo2013supporting}. With machine learning techniques, we can with less effort build prediction models that fit specific use cases and achieve high accuracy. Historically, Neural Networks (NN) based approaches have been used extensively. 
In this research, in addition to the Deep Learning (DL) models, we evaluate Decision Tree (DT) based ensemble methods and Genetic Programming (GP) to implement the audiovisual quality function to predict the perceived quality directly from the parameters extracted from the application and network layers. The machine learning based models capture the complex relationships between influence factors no matter if the dataset is generated for IPTV services or video-telephony in mind.

\section{Machine Learning Algorithms}
\label{sec:ml}

\subsection{Decision Tree Based Ensemble Methods}
Decision Trees (DT) are hierarchical data structures that can be used for classification and regression problems effectively using the divide-and-conquer strategy. A Decision Tree is composed of internal decision nodes where a test is applied to a given input and branches to a classification or regression value by the leaf nodes. The estimation process originates at the root node, traverses the decision nodes until a leaf node is hit \cite{alpaydin2014introduction} \cite{mushtaq2012empirical}.

The tree structure allows a fast discovery of nodes that cover an input. In a binary tree,
traversing each decision nodes exclude half of the cases ideally, if the tree is balanced. Due to fast convergence and ease of interpretation, they are sometimes preferred over more accurate methods \cite{alpaydin2014introduction}.

The estimation can be computed in a parametric model as well as a nonparametric model. In
parametric estimation, the model is built over the whole input space from the training data and a static tree structure is formed. Then the same model is used to make estimations as test data arrives. In the nonparametric approach, the tree structure is not static and, during the learning process, it grows as branches and leaves are added \cite{alpaydin2014introduction}.

Decision Trees have low bias and very high variance which bring over-fitting issues when they grow very deep. To reduce variance, Decision Tree-based ensemble methods have been developed. Random Forests (RF) are such an ensemble learning method for classification and regression that utilize several Decision Trees models to obtain a better prediction performance. During training, an array of Decision Trees formed and a randomly chosen subset of training data is used to train each tree. In a classification problem, the inputs are submitted to every tree in the RF in order to get a vote for a class. An RF model collects all votes and then picks the class with the highest number of votes. That behavior reduces the high variance issues we have mentioned above. However, since there is a tradeoff between bias and variance, RF classification introduces a small increase in bias while reducing variance. Overall it still provides significant improvements in terms of classification accuracy \cite{mushtaq2012empirical} \cite{breiman2001random}.

Rather than searching for a single superior model, researchers have noticed that combining many variations produce better results with a little extra effort. As we see for the Random Forests, ensemble learning models generate many classifiers and combine their results. This approach has been gaining a lot of interest recently. Two well-known methods of ensemble learning are boosting and bagging. In both methods, the learning algorithm combines the predictions of multiple base models \cite{domingos2012few} \cite{liaw2002classification} \cite{oza2005online}.

When we compare the Decision Trees based models with Bagging methods, each tree is constructed with a random variation of the training data set. The prediction is accomplished by the simple majority vote to improve stability and accuracy. This approach greatly reduces the variance as well as helping to avoid over-fitting issues, but it slightly increases the bias. Although it is usually applied to Decision Trees, it can be used with any type of method as well \cite{domingos2012few} \cite{liaw2002classification}.

\subsection{Symbolic Regression and Genetic Programming}
\label{section:GeneticProgramming}
Symbolic regression techniques aim to identify an underlying mathematical expression that best fits a dataset. It consists of simultaneously finding both the form of equations and their parameters. Symbolic Regression starts by forming an initial expression by randomly combining mathematical building blocks and then continue forming new equations by recombining previous equations using Genetic Programming (GP) \cite{schmidt2010symbolic}.

GP is a computation technique that enables us to find a solution to a problem without knowing the form of the solution in advance. It is based on the evolution of a population of computer programs where populations are transformed stochastically into new populations generation by generation \cite{poli2008field}.

GP discovers the performance of a program by running it, measuring its outcome, and then comparing the result to some objective. This comparison is called fitness. In the machine learning domain this would be equal to finding the ‘score’, ‘error’ or ‘loss’. In each generation, the programs that do well are marked to breed and then are used to produce new programs for the following generation. Crossover and mutation are the main genetic operations for creating new programs from existing ones. In the crossover, a child program is generated by joining randomly chosen parts from two selected programs from the previous generation. In a mutation, however, a child program is created from a single parent from the previous generation by randomly modifying a randomly selected segment \cite{poli2008field}.

GP usually utilizes trees in order to manipulate the programs. In the tree, function calls are represented by nodes and values associated with the functions are represented by leaves \cite{koza1992genetic}. GP programs combine multiple components in more advanced forms as well \cite{poli2008field}.

Similar to the ensemble methods we have seen in the previous section, initial GP populations are typically randomly generated as well. These initial populations are categorized as full, grow and ramped half-and-half depending on their depth \cite{poli2008field}.

Both full and grow methods limit the maximum depth of the initial individuals generated. They differ from each other with respect to the size and the shape of the trees generated. In the full method, trees are generated where all the leaves are at the same depth. In the grow method, trees are generated in various sizes and shapes. Ramped half-and-half method proposes a combination of both full and grow methods. In this approach, the full method is used to construct the half of the initial population and the grow method is used to construct the other half of the initial population \cite{poli2008field}.

GP selects the individuals probabilistically based on their fitness and then applies the genetic operations to them. This process causes better individuals to have likely more child programs than inferior individuals. Two common individual selection methods in GP are tournament selection and fitness proportionate selection \cite{poli2008field}.

\subsection{Deep Learning}

Deep Learning dates back to the 1940s and has been re-branded many times, reflecting the influence of different researchers and newer perspectives. it has only recently become “Deep Learning” \cite{bengio2015deep}.

A typical example of a Deep Learning model is the feedforward Deep Network or Multi-Layer perceptron (MLP) \cite{bengio2015deep}. A Multi-Layer Perceptron makes no assumptions about relationships among variables. In general, these models use three main layers: one input neurons layer that represents the input vector, one or more intermediator “hidden” layers and output neurons that represent the output vector. Nodes in each layer are linked to all nodes in adjoining layers. These links are used to forward signals from one neuron to the other \cite{mushtaq2012empirical} \cite{comrie1997comparing}.

Nonlinearities are represented in the network by the activation and transfer functions in each node. Each node handles a basic computation while its links enable an overall computation. The overall behavior of a Neural network is influenced by the number of layers, the number of neurons in each layer, how the neurons are linked and the weights associated with each link. The weight associated with each link defines how a first neuron influences the second neuron. During the training period, the weights are revised. With that approach, hidden layers capture the complexities in the data while the weights are adjusted in each iteration in order to obtain the lowest error in the output. The learning algorithm used most often is gradient descent backpropagation \cite{mushtaq2012empirical} \cite{bengio2015deep} \cite{comrie1997comparing}.

In the back propagation approach, during the forward phase, the input signal is propagated through the network layer by layer. In the output node, the error signal is computed and then this error signal is sent to the network in backward direction which is called the backward phase. During this backward phase, network parameters are modified in order to minimize the signal error \cite{du2009research}. Deep Learning methods can be used in regression problems as well as clustering and classification applications.

In Section \ref{sec:bitstreammodels} we will look into the details of these algorithms once again but this time from an implementation point of view and mention their specific configurations targeting regression usage. However, before that, we need to look at the INRS audiovisual quality dataset to see which information is available to us when attempting to build perceived quality estimation models.

\newcommand{\frigginfigurewidth}{2.4in}
\section{The INRS Audiovisual Quality Dataset}
\label{sec:dataset}

The Qualinet Multimedia Databases set v5.5 \cite{fliegel2014qualinet} provides a list of some publicly available audiovisual databases such as University of Plymouth \cite{goudarzi2010audiovisual}, TUM 1080p50 \cite{keimel2012tum}, VQEG \cite{pinson2012influence}, Made for Mobile \cite{robitza2012made} and VTT \cite{maki2013reduced} datasets. Additionally, the TVM and P.NAMS training and validation datasets \cite{garcia2014parametric} have led to ITU-T standards P.1201 \cite{itut2012P.1201} and G.1071 \cite{itut2015G.1071}.

Existing audiovisual datasets are invaluable for building perceived quality estimation models for one-way streaming based services. However, some of these datasets have some limitations in terms of real-time communication use cases, mainly because they have been generated with a different scope in mind. The University of Plymouth dataset is interesting with respect to real-time communications; however, video frame rates and the video resolution selected no longer reflect contemporary applications. In this research, we specifically target one-to-one real-time communications while exploring the encoding configurations -H.264 video codec, AMR-WB audio codec for speech signals, range of video frame rate, quantization and noise reduction parameters- and network impairments -range of packet loss rate for video and audio streams- often seen in contemporary video-telephony applications.

The INRS audiovisual quality dataset has been designed to span the most important compression and network distortion influence factors. These factors are typically video frame rate, quantization and filters, and network packet loss rate. We have chosen the range of these parameters for the H.264 video encoding as follows: 0, 0.1, 0.5, 1 and 5\% for network packet loss rate (PLR) in both video and audio streams, 10, 15, 20 and 25 FPS for video frame rate, 23, 27, 31 and 35 for quantization parameter (QP), and 0 and 999 for the noise reduction (NR) filter (Table \ref{tab:dataset_media}).

\begin{table}
\centering
\caption{Media compression parameters and network impairments.}
\label{tab:dataset_media}
\begin{tabular}{|l|l|l|} \hline
& Video & Audio \\ \hline
FPS & 10, 15, 20, 25 & Mono, 16kHz, 24 kbps \\ \hline
QP & 23, 27, 31, 35 & Mono, 16kHz, 24 kbps  \\ \hline
NR & 0, 999 & Mono, 16kHz, 24 kbps \\ \hline
PLR (\%) & 0, 0.1, 0.5, 1, 5 & 0, 0.1, 0.5, 1, 5 \\ \hline
\end{tabular}
\end{table}

The original 42 seconds long audiovisual raw source was encoded with the H.264/AVC video codec and the AMR-WB audio codec and then multiplexed into a 3gp container with the GStreamer open source multimedia framework \cite{teamgstreamer}.  Video streams were encoded with the constrained-baseline profile at a 720p progressive video resolution. Audio encoding settings were kept the same -mono channel, 16 kHz sample rate, and 24Kbps bit rate- for all audiovisual sequences. The multimedia framework, GStreamer, uses only the jitter buffer mechanism to smooth out the packet flow and has no forward error correction strategy. Hence in our dataset we assume that the packet loss figures reported are the residual packet losses.

An emulated network was used to transmit and record the audiovisual sequences. The audio and videos streams were captured with our GStreamer based custom developed software which enabled us to gather detailed RTCP statistics and report the exact network packet loss values for video and audio streams separately. This software is freely available to public access \cite{demirbilek2016githubGStreamertestbed} and detailed information on the technical implementation of this testbed and how each of these configuration settings is implemented is presented in \cite{demirbilek2016multimedia}. The Netem network emulator was deployed to produce network packet loss conditions. Network packet loss was only activated after the first second of the audio and video transmission. This allowed us to be sure that communications were initiated, especially for the video stream. get more realistic results. A custom video player was developed to collect the subjective scores \cite{demirbilek2016githubplayer}. Thirty observers have rated the overall audiovisual quality on the Absolute Category Rating (ACR) 5-level quality scale in a controlled environment. It is important to note that the required number of observers as well as post-processing of scores submitted by each observer used in this research are based on the latest ITU-T Recommendation P.913 \cite{itu913subjective}. However, evolution of the contribution of subjective assessment factors such as subjective bias, subject inaccuracy, stimulus scoring, personal and cultural traits on perception and consequently how subjective experiments should be conducted is a matter of ongoing research.

From our earlier work on the parametric models \cite{demirbilek2017parametricmodels}, we know that the change in the network packet loss rate has dramatic effects on the perceived quality. Moreover, video frame rate and quantization parameter have also a moderate influence on MOS value and model behavior. However, in the parametric version of the dataset, the values of these parameters are reported globally, for each file. We would expect that incorporating the influence of these parameters per video I and P frames, and per audio frames would improve the accuracy of the machine learning-based perceived quality prediction models. In the rest of this section, we will discuss the bitstream version of the INRS audiovisual quality dataset that we have created in order to gather packet loss rate and bit rate changes on the individual or group of frames. 

In light of these goals, we have used the FFmpeg multimedia system \cite{ffmpeg} to obtain the bit rates for audio and video streams, number of frames, and the duration of the streams.

Using the partial implementation of the ITU-T P.1201.2 audiovisual quality estimation tool at \cite{itut2012P.1201.2_tool}, we have computed the BitPerPixel and SceneComp parameters for video files in the INRS audiovisual quality dataset.

Additionally, we have reported the size and the loss percentage during the transmission of video I-frames individually as there is only one I frame every 10 seconds in the reference videos. We have however taken a different approach for the audio frames and video P frames. Instead of reporting each individual frame, we have grouped these frames on per second basis and reported the loss percentage in the count of audio frames and the loss percentage in the count and the average size of video P frames. During our initial experiments with machine learning based modeling, we have realized that reporting values for the first three video I-frame periods (i.e. the first 30 seconds) is sufficient for improving the model performance and the two remaining video I-frame periods would just make the model unnecessarily complex without any significant improvement in the model accuracy. The significance of the first 30 seconds is also in line with the analysis in \cite{demirbilek2016inrsquality} that the observers did not watch the videos until to the end during the subjective assessment.

The newly generated bitstream dataset contains base parameters (5 features) from the Table \ref{tab:dataset_media} and an additional 120 new features. The complete list of all features is given in the Table \ref{tab:parameters} and explanations regarding the meaning of their names will be given later.

\begin{table}
\centering
\caption{Bitstream Dataset features.}
\label{tab:parameters}
\begin{tabular}{|l|l|} \hline
Video Frame Rate & Noise Reduction \\ \hline
Quantization Parameter &  Video Packet Loss Rate \\ \hline
Audio Packet Loss Rate & Video Start PTS \\ \hline
Video Duration TS & Video Bit Rate \\ \hline
Video NB Frames & Audio Duration TS \\ \hline
Audio Duration & Audio Bit Rate \\ \hline
Audio NB Frames & Video Bits (Pixel*Frame) \\ \hline
iFrames Per Scene & Content Complexity \\ \hline
iFrame Count & pFrame Count \\ \hline
aFrame Count & iFrame Count Diff \\ \hline
pFrame Count Diff & aFrame Count Diff \\ \hline
\multicolumn{2}{|l|}{iFrame0 Size ... iFrame4 Size } \\ \hline
\multicolumn{2}{|l|}{iFrame0 Size Diff ... iFrame4 Size Diff } \\ \hline
\multicolumn{2}{|l|}{S0 pFrame Count Diff ... S30 pFrame Count Diff } \\ \hline
\multicolumn{2}{|l|}{S0 pFrame Mean Diff ... S30 pFrame Mean Diff } \\ \hline
\multicolumn{2}{|l|}{S0 aFrame Count Diff ... S30 aFrame Count Diff } \\ \hline
\end{tabular}
\end{table}

The models that use the bitstream version of the INRS audiovisual quality dataset would qualify as Reduced-Reference quality models since we have used the limited amount of information from original video files when calculating the effect of the packet loss on the audio frames and video I and P frames. The bitstream version of the INRS audiovisual quality dataset is publicly available at \cite{demirbilek2016githubINRSdataset}. More detailed information on the INRS audiovisual quality dataset regarding the selected video sequence, test methodology and comparison to the other publicly available datasets is given in \cite{demirbilek2017parametricmodels} and \cite{demirbilek2016inrsquality}. 

As recent developments show \cite{maki2013reduced}  \cite{raake2011ip} \cite{garcia2014parametric} \cite{belmudez2015audiovisual} \cite{gastaldo2013supporting} \cite{goudarzi2010audiovisual}, No-Reference and Reduced-Reference parametric models achieve high accuracy in estimating perceived quality with limited resources. In the next Section, we try to build and analyze similar models by estimating the audiovisual quality based on the INRS audiovisual quality dataset directly using Random Forest, Bagging, Genetic Programming and Deep Learning machine learning methods.

\section{Reduced-Reference Bitstream Audiovisual Quality Prediction Models}
\label{sec:bitstreammodels}

In earlier research \cite{demirbilek2017parametricmodels}, we have used the INRS audiovisual quality dataset for building several machine learning based perceived quality estimation models. We have built No-Reference parametric perceived quality estimation models based on the Random Forests, Bagging, Deep Learning and Genetic Programming methods. We have observed that all of the mentioned methods have achieved high accuracy in terms of RMSE and Pearson correlation. Random Forests and Bagging based models have shown a small edge over Deep Learning with respect to the accuracy they provide on the INRS dataset. Genetic Programming based models have fallen behind even though their accuracy was impressive as well. In this paper, we present the reduced-reference bitstream audiovisual quality prediction models developed based on the algorithms used in our previous work. 

Random Forests and Bagging ensemble models are based on decision trees. Genetic Programming models via Symbolic Regression generate models composed of mathematical equations. Deep Learning models are based on the multiple layers of Neural Networks. In Section \ref{sec:ml}, we have discussed in details these different models' structures as well as the theoretical foundations of these algorithms.

\subsection{Feature Selection}

In order to empirically test the increased number of the features selected, we have used the feature importance information provided by the Random Forests based algorithm. 

\begin{figure}[ht!]
\includegraphics[width=3.2in]{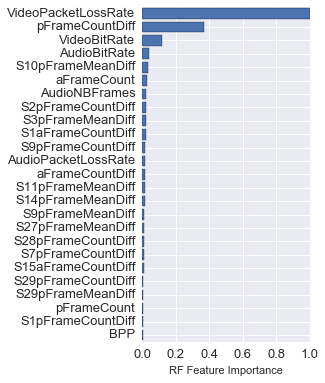}
\caption{Feature importance for the Random Forests model.}
\label{fig:RFFeatures}
\end{figure}

We have run leave-one-out cross-validation for every data-point in the bitstream dataset using the Random Forests based algorithm and have obtained the feature importance from each model. Afterwards, we have computed the overall feature importance by averaging the data we have collected from individual models. These averaged top 25 feature importance values are depicted in Figure \ref{fig:RFFeatures}. Here, many parameter names are intuitive and self explanatory. However, some other parameters might require additional discussion. 

Some of these parameters are given below.

\begin{itemize}

\item pFrameCount: Total count of P-frames in video stream.

\item pFrameCountDiff: Total loss rate in the count of P-frames.

\item S10pFrameMeanDiff: Loss rate in the average size of the P-frames for the 10th second.

\item S2pFrameCountDiff: Loss rate in the count of P-frames for the 2nd second.

\item aFrameCount: Total count of audio frames.

\item aFrameCountDiff: Total loss rate in the count of audio frames.

\item S1aFrameCountDiff: Loss rate in the count of audio frames for the 1st second.

\end{itemize}

Video packet loss rate influences the model performance more than other features by a great margin followed by the difference in the video P frame count difference. Additionally, we observe the video and audio bit rate, audio and video P frame count, audio packet loss rate, the difference in the audio frame count and BitPerPixel in the top 25 most influential features based on the Random Forests feature importance.  The rest of list consists of the effects of the packet loss rate on the individual per-second basis measurements. The order of seconds in the figure points to another quite important finding: Almost all most influential one second periods come right after the video I-frame transmission. This reflects the cumulative effect of packet loss and the importance of packet loss at the beginning of a period vs. at the middle or at the end of periods between two video I-frames. The appearance of the 9th seconds seems to contradict this conclusion. However, when we have carefully analyzed the bitstream dataset, for high packet loss rates, we have observed that for the transmitted video, some video I and P frames that are supposed to be in the 10th second period appear earlier in the 9th second period due to the way GStreamer framework handles high packet loss rates.

Another important finding is the absence of certain parameters in the 25 most influential features list. Quantization, video frame rate, and noise reduction have less influence on model behavior compared to other features. This behavior is due to the correlation of features: Random Forests feature selection prefers variables with more classes and when one of the correlated features is used, the importance of the other correlated features is reduced \cite{strobl2007bias}. We also observe the lack of the Scene-Complexity feature that we have computed in previous section. This behavior is easily explained by the fact that the INRS audiovisual quality dataset includes only one type of content.

These findings point to the following conclusions:

\begin{itemize}
  \item Packet loss rate is by a great margin the most dominant factor in modeling the perceived quality estimation.
  \item The effect of the packet loss is accumulated until the next video I frame and the packet loss at the beginning of the period is more important than at the middle or at the end of a period.
\end{itemize}

We have reorganized the dataset columns based on the findings from leave-one-out cross-validation by placing the most influential feature (VideoPacketLossRate) on the first column followed by the second most influential feature (pFrameCountDiff) in the second column and the rest of the columns sorted following the same logic. Sorting the dataset in this way has simplified our eventual Machine Learning models' source codes that we introduce in the following section. 

\subsection{Decision Trees Based Models} 

We have used the Python scikit-learn's implementation of Random Forests (RF) and Bagging (BG). We have generated Random Forest and Bagging models that used the bitstream dataset with 125 features. 

Initially, we have tuned our Random Forests and Bagging based model parameters using the 5 base parameters listed in Table \ref{tab:dataset_media}. Please note that packet loss rates are reported separately for audio and video streams. For the Bagging models, we have set the tree size to 166 and max\_features to 1.0 and max\_sample to 0.4 when looking for the best split and no limits on the tree depth. These parameters were 166 for tree size, 0.4 for max\_features and 0.8 for max\_sample to 0.8 for all Random Forests based models. 

To compare the effectiveness of models, we have generated 125 models using the Random Forest and 125 models using the Bagging algorithm where the 1st model used only the most influential feature, the 2nd model used two most influential features... and the 125th model used all of the features in the bitstream dataset. We have run each model 10 times using the 10-fold cross-validation and have reported statistical metrics averaged over these 10 runs. 

\begin{table}
\centering
\caption{Keras Deep Learning configurations.}
\label{tab:bitstreamKeras}
\begin{tabular}{|l||l|} \hline
Variable & Value \\ \hline
Model Type & Sequential  \\ \hline
Neural network Layers & Dense  \\ \hline
Activation Function & Tanh and Softplus  \\ \hline
Initialization Function & Uniform  \\ \hline
Optimizer & Adadelta  \\ \hline
Loss Function & MSE  \\ \hline
Batch Size & 4  \\ \hline
Epoch Size & 440  \\ \hline
\end{tabular}
\end{table}

\begin{table}
\centering
\caption{gplearn Genetic Programming configurations.}
\label{tab:bitstreamgplearn}
\begin{tabular}{|l||l|} \hline
Variable & Value \\ \hline
Model Type & Symbolic Regression  \\ \hline
Population Size  & 5000  \\ \hline
Generations & 200  \\ \hline
Tournament Size & 20  \\ \hline
Stopping Criteria & 0.0  \\ \hline
Init Method & full  \\ \hline
Transformer & True  \\ \hline
Comparison & True  \\ \hline
Trigonometric & False  \\ \hline
Metric & RMSE  \\ \hline
Parsimony Coefficient & 0.001 \\ \hline
P Crossover & 0.9 \\ \hline
P Subtree Mutation & 0.01 \\ \hline
P Hoist Mutation & 0.01 \\ \hline
P Point Mutation & 0.01 \\ \hline
P Point Replace & 0.05 \\ \hline
Max Samples & 0.8 \\ \hline
Jobs & 1 \\ \hline
Random State & None \\ \hline
\end{tabular}
\end{table}

\subsection{Deep Learning Based Models}

We have generated the Deep Learning (DL) models using the Keras Deep Learning library that run on top of the Theano library. Similar to Decision Trees based models, we have initially tuned the DL model parameters using the 5 features listed in Table \ref{tab:dataset_media}. Eventually, we have obtained the DL model configuration listed in Table \ref{tab:bitstreamKeras} using only one hidden layer and have generated 125 models based on the increased number of features. Note that eventhough our final DL model configuration does not consist of multiple hidden layers, we kept the naming as is as multiple hidden layers did not provide a superior performance. For the DL models, we have run each model 10 times using the 4-fold cross validation and have reported statistical metrics averaged over these 10 runs.

\subsection{Genetic Programming Based Models}

We have used the gplearn library to implement genetic Programming (GP) based models. The gplearn library is similar to scikit-learn fit/predict API and works with the existing scikit-learn pipeline. We have used the SymbolicRegressor with the configurations listed in the Table \ref{tab:bitstreamgplearn}. Similar to the DL models, we have run each model 10 times using the 4-Fold cross validation. 

\begin{table*}
\centering
\caption{Best performing Models for RF, BG, DL and GP algorithms.}
\label{tab:maxminvalues}
\begin{tabular}{|l||l|l||l|l||l|l|} \hline
Algorithm & Max Pearson Cor. & Features & Min RMSE & Features & Min RMSE$^*$ & Features \\ \hline
Random Forests & 0.9328 & 18 & 0.3514 & 18 & 0.1478 & 15 \\ \hline
Bagging & 0.9332 & 20 & 0.3547 & 11 & 0.1565 & 8 \\ \hline
Deep Learning & 0.9130 & 7 & 0.3970 & 7 & 0.1711 & 16 \\ \hline
Genetic Programming & 0.8928 & 7 & 0.4364 & 7 & 0.1787 & 7 \\ \hline
\end{tabular}
\end{table*}

\begin{figure*}[ht!]
\includegraphics[width=3.6in]{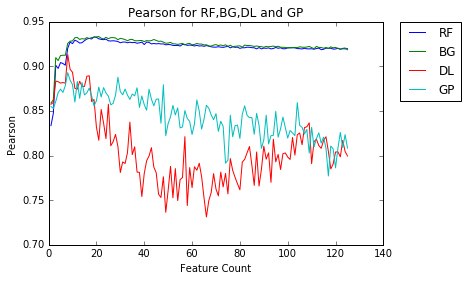}
\includegraphics[width=3.6in]{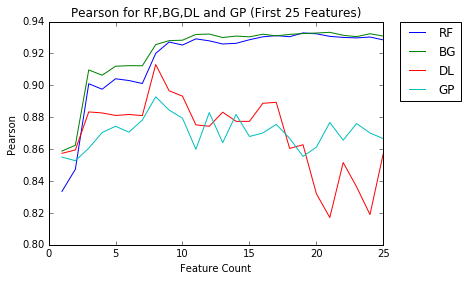}
\includegraphics[width=3.6in]{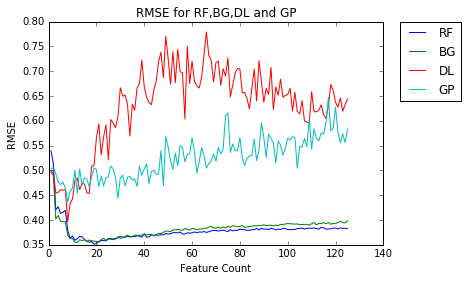}
\includegraphics[width=3.6in]{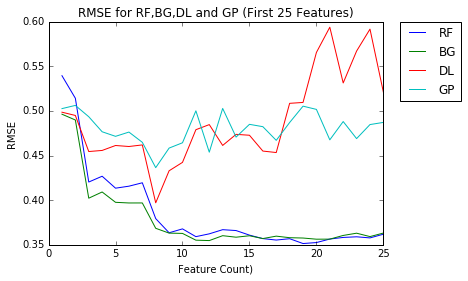}
\includegraphics[width=3.6in]{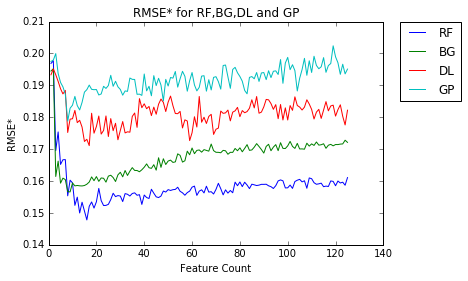}
\includegraphics[width=3.6in]{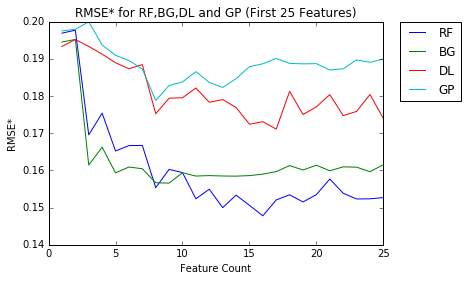}
\caption{Pearson Correlation, RMSE and RMSE$^*$ for RF, BG, DL and GP Models.}
\label{fig:all}
\end{figure*}

\subsection{Results}

Figure \ref{fig:all} depicts the Pearson correlation coefficient, RMSE and RMSE$^*$ values obtained for each algorithm. The values for 125 models are shown on the left side while a closer look at the metrics for models generated up to the top 25 features is shown on the right side. Table \ref{tab:maxminvalues} shows the maximum Pearson correlation, minimum RMSE and RMSE$^*$ values and the number of features used for each of these models with each algorithm.

Random Forests and Bagging based models perform best using around the top 20 features while Deep Learning and Genetic Programming based models reach to their maximum with the 7 top features in terms of Pearson Correlation coefficient. Decision Trees based models more or less perform similarly with increased number of features. However, both Deep Learning and Genetic Programming based models suffer significantly from increased number of features in terms of all metrics we have used. Overall, looking at the Pearson Correlation and RMSE figures, we can conclude that Decision Trees based models make more accurate estimations of MOS values followed by Genetic Programming and then Deep Learning models. 

RMSE$^*$ graphs show the difference between each algorithm more clearly. Although all models perform very similarly with the top 2 features, increasing the number of the features sets them apart. Random Forests based models overall perform best followed by the Bagging, Deep Learning and then Genetic Programming models respectively.

In light of this information, let us revisit Figure \ref{fig:RFFeatures}. Both Deep Learning and Genetic Programming based models perform best with the top 7 features. This is in line with our expectations that Deep Learning based models do not perform well with an increased number of features and require prior feature engineering. Decision Trees based models are capable of integrating per-second basis reduced-reference parameters and achieve to their highest potential with about the top 20 features. 

\begin{figure*}[ht!]
\includegraphics[width=3.6in]{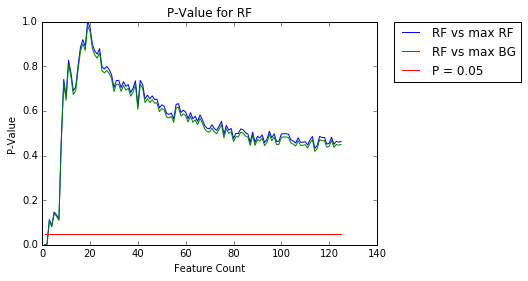}
\includegraphics[width=3.6in]{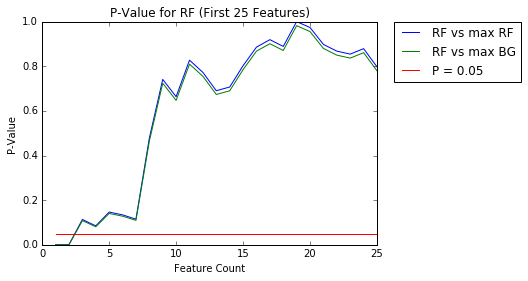}
\includegraphics[width=3.6in]{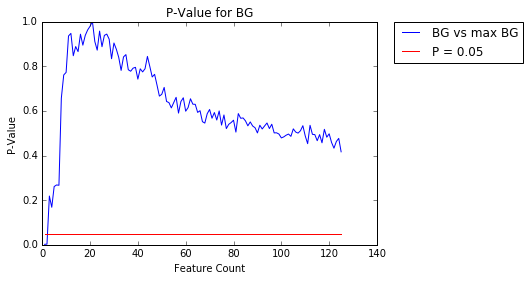}
\includegraphics[width=3.6in]{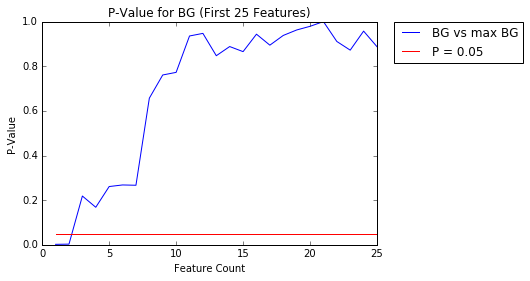}
\includegraphics[width=3.6in]{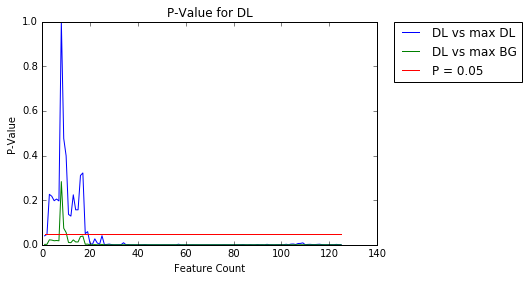}
\includegraphics[width=3.6in]{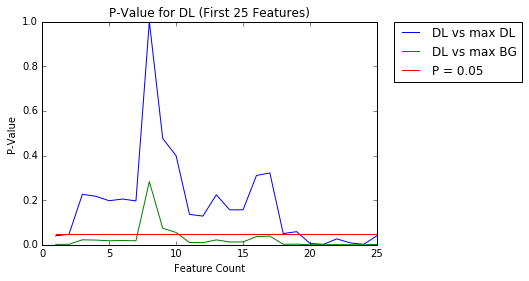}
\includegraphics[width=3.6in]{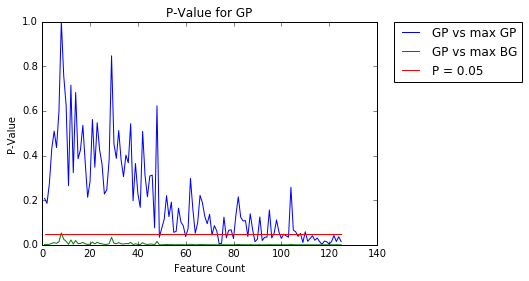}
\includegraphics[width=3.6in]{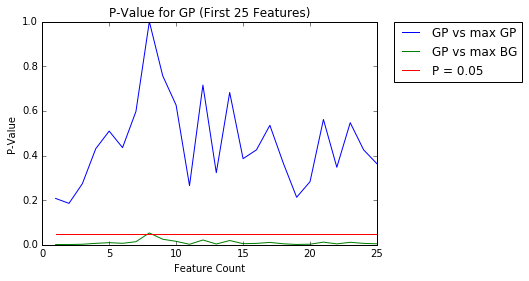}

\caption{Statistical Significance Test for RF, BG, DL and GP Models. The graphs on the left show the evolution of the P-value with increasing number of features for each algorithm and the graphs on the right focus on the first 25 features.}
\label{fig:stats}
\end{figure*}

Since we have shuffled the order of the rows in the dataset each time before running the models, the results that we have obtained had different values depending on the training and tests sets selected and algorithm utilized. To reduce the variation in metrics, we have run each model consecutively 10 times (with the entire dataset shuffled before each run) and have taken the average of the measured statistical metrics in these 10 runs for every reported performance indicators (Figure \ref{fig:all}). Random Forest and Bagging based models have not only achieved the highest accuracy in terms of Pearson correlation and RMSE but are also more precise and have less variation in measured metrics.

Another important and related dimension of the modeling is finding if two performance metrics are statistically significantly different. In Section \ref{subsec:Metrics} we have seen that ITU-T P.1401 \cite{itut2012P.1401} recommends computing the significance of the difference between correlation coefficients using the  Fisher-z transformation. In order to achieve that, we have conducted two statistical test for each individual model. First, for each algorithm, we have found the best performing model in terms of Pearson correlation coefficient and then we have compared the other 124 models based on the same algorithm with that selected best performing model. Second, we have found that the globally best performing model in terms of Pearson correlation is the Bagging model using the top 20 features, and we have compared it with the all of other models and reported the P-values \cite{sellke2001calibration}. These computations are depicted in the Figure \ref{fig:stats} where the graphs on the left show the evolution of the P-value with increasing number of features for each algorithm while graphs on the right focus on the first 25 features. The significance level $\alpha$ for a given hypothesis test is a value for which a P-value less than or equal to $\alpha$ is considered statistically significant \cite{sellke2001calibration}. Typical values for $\alpha$ are 0.1, 0.05 and 0.01. In the figures, we have shown the standard cutoff value of P at 0.05. When computing these values, we have used the sample size as 160 even though the Pearson Correlation coefficients values reported are the average of 10 runs. In general, the difference between two correlation coefficient tend to be statistically more significant with the increased sample size. Therefore, the results shown in the Figure \ref{fig:stats} should be considered with the given sample size in mind.

In our experiments, we have observed that Decision Trees based models perform better in accuracy and precision and also required less effort to generate. Random Forests algorithms also provide feature importance ranking which was invaluable for feature selection.

\section{Discussion}

The standardized models for IPTV \cite{itut2015G.1071} \cite{itut2012P.1201} and video telephony \cite{itut2012G.1070} aim to provide a model for as many use cases as possible. They typically utilize a very small subset of the features available such as packet loss percentage, frame rate and compression rate, and content complexity for bitstream models. In our earlier work \cite{demirbilek2017parametricmodels} we have discussed these standardized models in details. In this research, we have generated audiovisual perceived quality estimation models using both the traditional small set of features used in standardized models and the correlated data that we have extracted from the videos included in the publicly available dataset. One of the main difference between these standard models and the approach we have taken here is taking advantage of the correlated data. Based on the results that we have obtained in this research as well as results we have obtained in our earlier work, we know that extracting additional correlated data from the dataset helps us to generate more accurate models. However, this correlated data depends on various factors such as the audio and video codec used, the tool used to extract the correlated data and number of features. For similar research, we recommend following the approach we have taken here rather than pinpointing specific parameters. In our previous work \cite{demirbilek2017parametricmodels}, we have also tested the Random Forest based models on the University of Plymouth \cite{goudarzi2010audiovisual}, TUM 1080p50 \cite{keimel2012tum} and VQEG datasets \cite{pinson2012influence} and achieved the comparable results with the original studies on these datasets.

In this research, we have chosen media and channel encoding parameters that reflects contemporary real- time configurations for video frame rate, video quantization, noise reduction parameters and network packet loss rate. Therefore, the models we have developed in our research are comparable to the ITU-T G.1070 \cite{itut2012G.1070}, G.1071 \cite{itut2015G.1071} and P.1201 \cite{itut2012P.1201} standard models. Running first two models on our dataset was not possible as these models have limited coverage and do not support the AMR-WB audio codec combined with H.264 video codec at 720p video resolution. The ITU-T G.1071 model supports AMR codec only in low resolutions up to HVGA (480 x 320) resolution. The ITU-T G.1070 model provides model constants, m, only for 2.1 and 4.2-inch video display sizes and also does not provide speech coding-distortion and packet-loss-robustness constants for AMR codec. Comparison with the ITU-T P.1201 model is not given as it does not support the AMR-WB codec.

\section{Conclusion}
\label{sec:conclusion}

We have developed bitstream perceived quality estimation models based on the Random Forests, Bagging, Deep Learning and Genetic Programming methods using the bitstream version of the INRS audiovisual quality dataset which includes media compression and network degradations typically seen in real-time communications.

During the model generation, instead of selecting a specific subset of features, we have taken an empirical approach and have generated models varying from very simple to the most complex depending on the number of features used from the quality dataset to find out how each algorithm perform with the correlated data. 

The bitstream version of the INRS audiovisual quality dataset consists of 125 features. In order to rank the features, we have initially run leave-one-out cross-validation for every data-point in the bitstream dataset using the Random Forests based algorithm and have obtained the feature importance from each model. Afterward, we have computed the overall feature importance by averaging the data we have collected from individual models. Then, using this feature importance information, we have generated 125 models for each algorithm where the 1st model used only the most influential feature, the 2nd model used the two most influential features... and the 125th model used all of the features in the bitstream dataset.

Random Forests and Bagging based models perform best using around the top 20 features while Deep Learning and Genetic Programming based models reach to their maximum with the 7 top features in terms of RMSE values and Pearson Correlation coefficient. We have additionally computed the RMSE$^*$ values and have computed the significance of the difference between correlation coefficients using the Fisher-z transformation. RMSE$^*$ values showed that, although all models perform very similarly with the top 2 features, increasing the number of the features makes a difference. Random Forests based models overall perform best followed by the Bagging, Deep Learning and then Genetic Programming models respectively. It is also important to note that, generating decision trees based models require significantly less effort and less time due to a low number of configuration parameters they require. On the other hand, Genetic Programming models required the most effort to tune, the most computing resources to deploy and the longest training time. 

Overall we conclude that computing the bitstream information is worth the effort it takes to generate and helps to build more accurate models for real-time communications. However, it is useful only for the deployment of the right algorithms with the carefully selected subset of the features. Our studies have proved that Decision Trees based algorithms are well suited to the No-Reference parametric models as well as to the Reduced-Reference bitstream models.

\ifCLASSOPTIONcaptionsoff
  \newpage
\fi



\bibliographystyle{IEEEtran}
\bibliography{IEEEabrv,barebiblio.bib}
\end{document}